\documentclass[conference]{IEEEtran}

\usepackage{cite}
\usepackage{amsmath,amssymb,amsfonts}
\usepackage{booktabs}
\usepackage{graphicx}
\usepackage{multirow}
\usepackage{textcomp}
\usepackage{url}

\def\BibTeX{{\rm B\kern-.05em{\sc i\kern-.025em b}\kern-.08em
    T\kern-.1667em\lower.7ex\hbox{E}\kern-.125emX}}

\title{VoxTubeS: Distributable Speaker-Anonymized Synthetic Speech Corpora and Their Analysis}

\author{\IEEEauthorblockN{Zhe Zhang}
\IEEEauthorblockA{\textit{National Institute of Informatics} \\
Tokyo, Japan \\
zhe@nii.ac.jp}
\and
\IEEEauthorblockN{Yexin Lu}
\IEEEauthorblockA{\textit{University of Science and Technology of China} \\
Hefei, China \\
yxlu0102@mail.ustc.edu.cn}
\and
\IEEEauthorblockN{Junichi Yamagishi}
\IEEEauthorblockA{\textit{National Institute of Informatics} \\
Tokyo, Japan \\
jyamagis@nii.ac.jp}
}

\begin{document}
\maketitle

\begin{abstract}
Large speech corpora support research, but recordings can expose speaker identity because voice remains a recognizable biometric. Meanwhile, speech data derived from media can be difficult to redistribute reliably. We present \emph{VoxTubeS}, a family of speaker-anonymized synthetic speech corpora designed for redistribution, comprising three method families and seven variants derived from the VoxTube corpus, which is distributed under CC BY-NC-SA 4.0, using 1.29M quality-filtered English utterances from 1,511 speakers. The synthesis methods span voice conversion, latent-space anonymization, and controllable text-to-speech. We evaluate VoxTubeS using utterance-level unlinkability, conversation-level linkability and singling-out, downstream speaker verification, linguistic consistency, speaker diversity, and fairness metrics for gender and accents.  Our comprehensive analysis exposes a complex trade-off: stronger identity suppression often reduces linkability but sacrifices utility and population diversity, whereas speaker consistency training improves both utterance- and conversation-level privacy while retaining comparable utility and a broader speaker space. Fairness varies independently of aggregate performance. No method dominates; VoxTubeS therefore treats corpus construction as a choice among operating points that balances privacy, utility, diversity, fairness, and responsible redistribution under the source license.
\end{abstract}

\begin{IEEEkeywords}
speaker anonymization, synthetic speech corpora, voice privacy, fairness
\end{IEEEkeywords}

\section{Introduction}

Large public speech corpora have been central to progress in speaker recognition, diarization, speech recognition, and speech generation \cite{roger2022speechsurvey}. They also create a persistent privacy problem: speech is not merely acoustic content, but a biometric signal that can reveal the identity of the speaker, demographic attributes, the recording context, the health state, and the social environment. For speaker recognition research, the conflict between privacy and utility is structural: training requires speaker-discriminative variation to learn useful embeddings, but such variation can make the released data linkable, searchable, and reusable in ways that the speakers did not anticipate.

Speaker anonymization and speech synthesis offer a possible route toward reusable privacy-aware speech resources. The VoicePrivacy initiative has progressively strengthened this evaluation setting through complementary measures of privacy, intelligibility, naturalness, and distinctiveness \cite{tomashenko2022voiceprivacy2020,panariello2024voiceprivacy2022,tomashenko2024voiceprivacyplan,tomashenko2026thirdvoiceprivacy}. However, strong utterance-level anonymization does not necessarily produce a useful training corpus. We therefore ask whether speaker-anonymized speech can reduce identity exposure while preserving downstream utility, linguistic fidelity, speaker-space structure, and fair subgroup performance.

This paper studies these corpus-level questions using VoxTube, a large-scale in-the-wild corpus collected from CC BY-licensed YouTube videos and distributed by under CC BY-NC-SA 4.0 \cite{yakovlev2023voxtube}. However, licensing permission alone does not address biometric exposure. We therefore treat privacy protection and responsible release as complementary requirements. This distinction matters because access to biometric corpora can be fragile: for example, the official VoxCeleb website no longer distributes source media or identifying metadata \cite{voxcelebAvailability}. 

\begin{figure}[t]
\centering
\includegraphics[width=\linewidth]{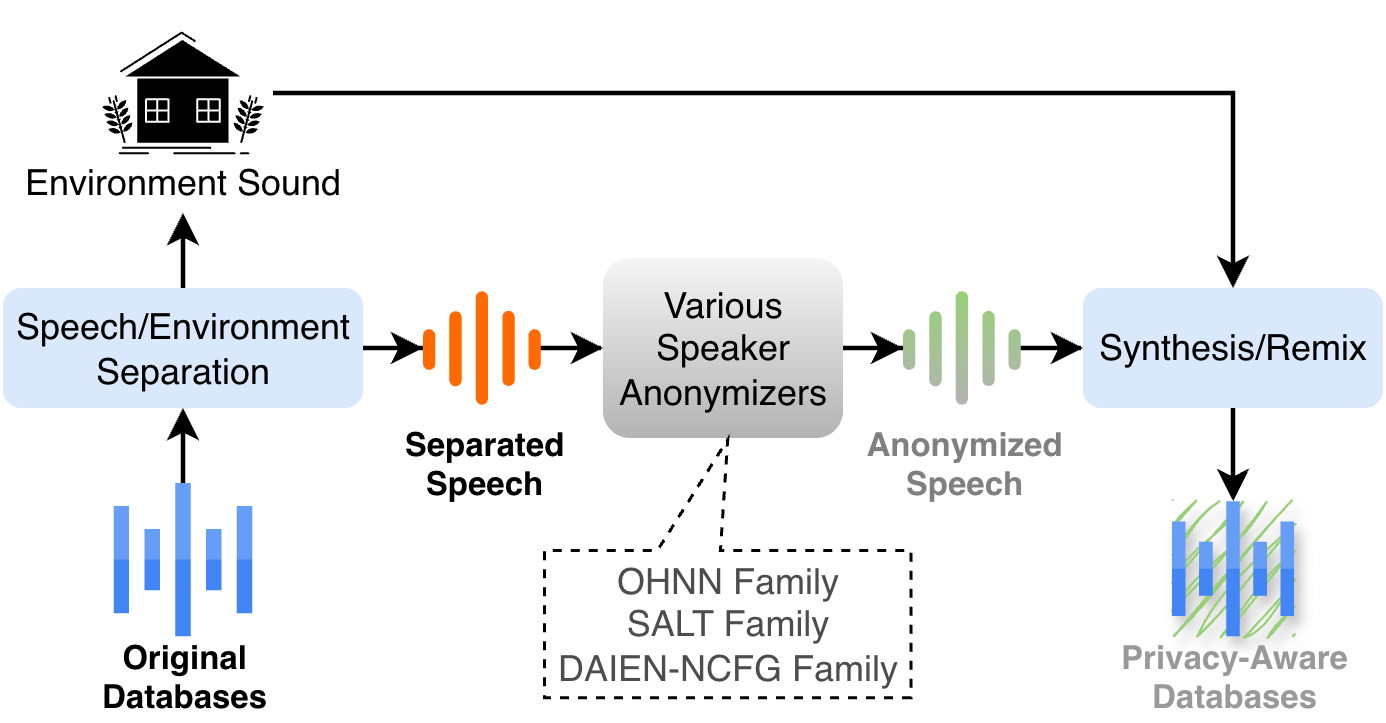}
\caption{VoxTubeS construction pipeline. OHNN and SALT anonymize an estimated speech component and then remix the estimated environment component, whereas DAIEN-NCFG synthesizes speech and environmental sound jointly from text and separate prompts. Environmental sound is retained to preserve the in-the-wild acoustic context.}
\label{fig:overview}
\end{figure}

As showin in Fig.~\ref{fig:overview}, we construct seven utterance-aligned variants using various synthetic methods from the same source: two orthogonal Householder neural network (OHNN) \cite{miao2023ohnn} variants, two SALT \cite{lv2023salt} variants, and three DAIEN-NCFG variants. \emph{OHNN-HiFiGAN} uses the OHNN speaker anonymization and HiFiGAN vocoder \cite{kong2020hifigan}; \emph{OHNN-BigVGAN-SC} also uses the same OHNN but adopts BigVGAN vocoder \cite{lee2023bigvgan} trained with a speaker-consistency (SC) objective. \emph{SALT-k4} and \emph{SALT-k8} use latent-space anonymization, where $k$ is the number of nearest target latent vectors averaged by frame-level k-nearest-neighbor (kNN) regression. \emph{DAIEN-NCFG($\gamma$)} denotes our negative classifier-free-guidance (NCFG) \cite{ho2022classifierfree} use of DAIEN-TTS \cite{lu2026daien}, where $\gamma$ is the speaker guidance weight and is set below zero to suppress the prompted speaker.

The OHNN and SALT pipelines can process the multilingual VoxTube inventory, whereas DAIEN-NCFG is limited to English by the underlying TTS model and text-conditioning pipeline. We therefore use the quality-filtered English subset for the paired comparison of all seven variants.

We call the resulting resource framework \emph{VoxTubeS}, where ``\emph{S}’’ denotes both synthetic and the plurality of variants. The seven variants are evaluated independently rather than combined into a single training corpus, providing distinct operating points for comparing privacy, utility, diversity, and fairness. We first describe the construction of the seven derivatives and then evaluate them as training data for automatic speaker verification (ASV), together with their privacy, linguistic consistency, speaker diversity, and subgroup behavior. This controlled million-utterance study shows why corpus-level anonymization cannot be reduced to one privacy score: greater identity suppression can destroy useful population structure, while high downstream utility can coexist with residual identity cues that still enable speaker linkage across utterances.

Our contributions are threefold. First, we construct and release VoxTubeS\footnote{\url{https://zenodo.org/records/22699494}}\footnote{\url{https://huggingface.co/datasets/nii-yamagishilab/VoxTubeS}} as a family of utterance-aligned synthetic derivatives designed for redistribution, including multilingual versions of OHNN and SALT, and a common English protocol for comparison with DAIEN-TTS. Second, we introduce two corpus-construction adaptations: OHNN-BigVGAN-SC for preserving population structure and DAIEN-NCFG as a controllable TTS-based identity-suppression method. Third, we train separate speaker-verification models on each variant and jointly evaluate corpus-level privacy, downstream utility, linguistic consistency, speaker diversity, and subgroup fairness. VoxTubeS listening samples are available on the demo page\footnote{\url{https://nii-yamagishilab.github.io/voxtubes-demo-pages/}}.

\section{VoxTubeS Corpus Construction}

The original VoxTube corpus was collected from CC BY-licensed videos and redistributed under CC BY-NC-SA 4.0 \cite{yakovlev2023voxtube}; this permits our non-commercial redistribution, and VoxTubeS inherits the same license. In the following, we describe how we constructed the VoxTubeS corpora shown in Table~\ref{tab:corpora}.

\subsection{English Filtering Protocol}
Although VoxTube covers many languages, the DAIEN-TTS model and text-conditioning pipeline used are limited to English, which motivates the common English protocol used in this study.
VoxTube does not provide transcripts. We therefore transcribed the subset labeled as English in the VoxTube metadata using Whisper large-v3 \cite{radford2023whisper}. Manual inspection of the resulting transcripts revealed that this subset still contained non-English speech, transcripts with insufficient linguistic content, repetitive transcripts, and a DAIEN-TTS preprocessing failure affecting transcripts ending in non-ASCII characters. The filtering protocol is therefore necessary both to ensure a valid comparison among English-only synthesis systems and to produce a higher-quality release manifest.

The initial subset contains 1,334,157 utterances. We apply a language-identification confidence threshold of 0.9, require at least four words and 20 alphabetic characters to exclude uninformative short transcripts, remove repetitive transcripts, and exclude videos identified as non-English. This filtering process removes 41,990 utterances while retaining all 1,511 speakers. The resulting filtered English subset contains 1,292,167 utterances, with a 9:1 train/development split.

\subsection{Speaker-Anonymized Corpora with Environmental Sound} 

We generate speaker-anonymized corpora from the filtered authentic VoxTube English corpus while preserving environmental sound, based on the original train/development split. This paired construction holds linguistic content, speaker inventory, and split information across the different variants.

For the OHNN and SALT pipelines, we first estimate speech and environment components from each source recording. Following the background-restoration strategy adopted for the SynVox2 corpus \cite{miao2024synvox2}, only the speech component is anonymized; the estimated environment component is then remixed with the generated waveform to preserve the in-the-wild acoustic variation. DAIEN-NCFG is the exception because DAIEN-TTS already contains separate speech/environment prompting and environment-aware synthesis, so it produces the final mixture within its own pipeline.

\begin{table}[tbp]
\caption{Authentic and generated VoxTube corpora.}
\centering
\label{tab:corpora}
\footnotesize
\begin{tabular}{@{}p{0.12\linewidth}p{0.31\linewidth}p{0.47\linewidth}@{}}
\toprule
Paradigm & Variant & Configuration \\
\midrule

Authentic
& Auth
& Original VoxTube English subset \\

\midrule

\multirow{2}{*}{\shortstack[l]{Embedding\\Transform}}
& OHNN-HiFiGAN
& HiFiGAN vocoder \\
& OHNN-BigVGAN-SC
& BigVGAN vocoder with SC training \\

\midrule

\multirow{2}{*}{\shortstack[l]{Latent\\Conversion}}
& SALT-k4
& Four-speaker mixture; $k=4$ \\
& SALT-k8
& Eight-speaker mixture; $k=8$ \\

\midrule

\multirow{3}{*}{\shortstack[l]{Prompted\\TTS}}
& DAIEN-NCFG($-1.0$)
& Speaker-prompt weight $\gamma=-1.0$ \\
& DAIEN-NCFG($-0.75$)
& Speaker-prompt weight $\gamma=-0.75$ \\
& DAIEN-NCFG($-0.5$)
& Speaker-prompt weight $\gamma=-0.5$ \\

\bottomrule
\end{tabular}
\end{table}

Details of each variant are as follows.
\paragraph{\textbf{OHNN-based pipelines}}
Two OHNN-based pipelines, OHNN-HiFiGAN and OHNN-BigVGAN-SC, are used. Both use the same OHNN transform from each authentic speaker embedding to a pseudo-speaker embedding. The former follows the recipe for creating the SynVox2 corpus \cite{miao2024synvox2} with HiFiGAN. The latter replaces HiFiGAN with BigVGAN, whose periodic inductive bias and anti-aliased architecture are intended to improve waveform fidelity \cite{lee2023bigvgan}. During BigVGAN training, we introduce the SC objective using embeddings from an ECAPA-TDNN speaker encoder \cite{desplanques2020ecapa}, encouraging the vocoder to preserve the speaker characteristics represented by the input mel spectrogram and speaker embedding.

\paragraph{\textbf{SALT-based pipelines}}
SALT-based pipelines use WavLM-based \cite{chen2022wavlm} voice conversion. SALT constructs a target latent pool by sampling and weighting reference speakers, blends each source frame with the target features, and synthesizes the transformed sequence using HiFiGAN \cite{lv2023salt,kong2020hifigan}. 
Our generator samples reference speakers from the LibriSpeech speaker pack \cite{panayotov2015librispeech} and uses their weighted mixture as an anonymized speaker. SALT-k4 uses a four-speaker mixture with kNN regression width $k=4$, whereas SALT-k8 uses an eight-speaker mixture with $k=8$.

\paragraph{\textbf{DAIEN-TTS-based pipelines}}
We adopt DAIEN-TTS \cite{lu2026daien} as a TTS-based speaker anonymization method. DAIEN-TTS is designed to synthesize environment-aware speech from text using separate speaker and environment audio prompts. For anonymization purposes, we set its speaker-prompt guidance weight $\gamma$ below zero. A positive speaker-prompt guidance weight strengthens agreement with the prompted speaker, whereas a negative value moves generation away from that speaker condition and suppresses prompted identity. We construct three aligned corpora denoted DAIEN-NCFG($-1.0$), DAIEN-NCFG($-0.75$), and DAIEN-NCFG($-0.5$).

\section{Analysis of the VoxTubeS Corpora}

We analyze the constructed VoxTubeS corpora along six evaluation axes, extending the metrics commonly used in the VoicePrivacy Challenge \cite{tomashenko2022voiceprivacy2020,panariello2024voiceprivacy2022,tomashenko2026thirdvoiceprivacy}.

\subsection{Utterance-Level Unlinkability}
Unlinkability is quantified as the ASV equal-error rate (EER) in a setting where original recordings are used for enrollment and anonymized recordings are used for testing, relying on pretrained ECAPA-TDNN speaker verifiers \cite{desplanques2020ecapa,ravanelli2024speechbrain}. This setup mirrors the behavior of a privacy attacker’s model that attempts to recognize speakers and thus reflects the effectiveness of the privacy protection. The verifier is trained on the VoxCeleb dataset \cite{nagrani2017voxceleb}.

\subsection{Conversation-Level Privacy}
Building on legally grounded definitions of singling-out and linkability concepts, as recently operationalized for voice anonymization \cite{vauquier2025legally} and by related risk assessment \cite{giomi2023anonymeter}, we compute  conversation-level privacy metrics over lengths $L \in \{1,3,30\}$ and populations $N \in \{20,100,1000\}$. Authentic enrollment embeddings are compared with mean anonymized-conversation embeddings. Here, the linkability is top-1 speaker matching; singling-out records whether an enrollment embedding isolates exactly one test conversation above a calibration-derived threshold. Five runs and ten folds are averaged.

\subsection{Utility Metrics}
We compute three types of measures as utility metrics to show the usefulness of the audio in the constructed corpora. 

The first metric is the performance of ASV models trained separately on each VoxTubeS variant. We train an ECAPA-TDNN model using each VoxTubeS variant and evaluate it on the VoxCeleb1-O trials \cite{nagrani2017voxceleb} with EER, reported as a percentage. Each model uses 80-bin filterbanks, 192-dimensional embeddings, additive angular-margin loss \cite{deng2019arcface}, 3-s random fragments, speed, reverberation and noise augmentation, and 16 epochs for training. Lower EER indicates better utility.

The second utility metric assesses linguistic consistency before and after anonymization, quantified by pseudo-word error rate (pWER). We employ Whisper large-v3 \cite{radford2023whisper} and compare transcriptions of anonymized audio with the original audio on 29,109 development utterances. This measure is not equivalent to WER computed against ground-truth transcriptions.

The third set of metrics quantifies the speaker diversity present in the VoxTubeS corpora, a crucial aspect for speaker-focused downstream applications. To this end, we represent the population geometry using a single centroid $\mathbf{c}_i$ for each speaker. The mean pairwise cosine distance is defined as
\begin{equation}
\bar{d}=\frac{2}{S(S-1)}\sum_{i<j}
\left(1-\cos(\mathbf{c}_i,\mathbf{c}_j)\right),
\end{equation}
where larger values reflect bigger separation between speakers. 
Given the covariance matrix eigenvalues $\lambda_j$ of speaker centroids, we set $p_j=\lambda_j/\sum_k\lambda_k$. The effective rank is then $r_{\mathrm{eff}}=\exp\!\left(-\sum_j p_j\log p_j\right),$
which shows the number of embedding directions that have non-negligible variance \cite{roy2007effectiverank}. High values correspond to widely distributed speaker spaces.

\subsection{Subgroup Fairness}
The last metric assesses the fairness of the downstream ASV models trained on VoxTubeS with respect to gender and accents. We use female/male VoxCeleb2 trials \cite{chung2018voxceleb2} and ten sets of accent-specific VoxAccent trials \cite{estevez2023fairness}. We then compute the subgroup discrepancies in false acceptance rate (FAR) and false rejection rate (FRR), summarized by the fairness discrepancy rate (FDR) \cite{leschanowsky2024allbias}, defined $1-[0.95\Delta_{\mathrm{FAR}}+0.05\Delta_{\mathrm{FRR}}]$, where each $\Delta$ is the difference between the maximum and minimum subgroup error rates. FDR is evaluated at the pooled-EER threshold. Higher FDR indicates smaller gaps among subgroups and better fairness.

\section{Results}
Table~\ref{tab:main} summarizes the evaluation results. The following subsections discuss the main findings.

\subsection{Privacy--Utility--Diversity Trade-offs}

\begin{figure}[t]
\centering
\includegraphics[width=\linewidth]{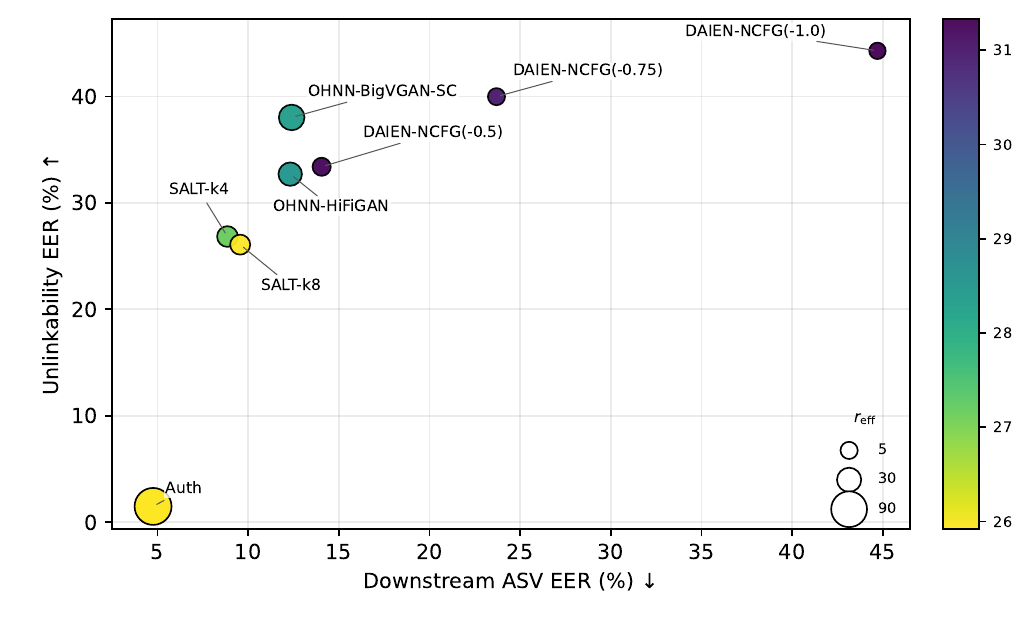}
\textbf{(a)} Utterance-level unlinkability
\includegraphics[width=\linewidth]{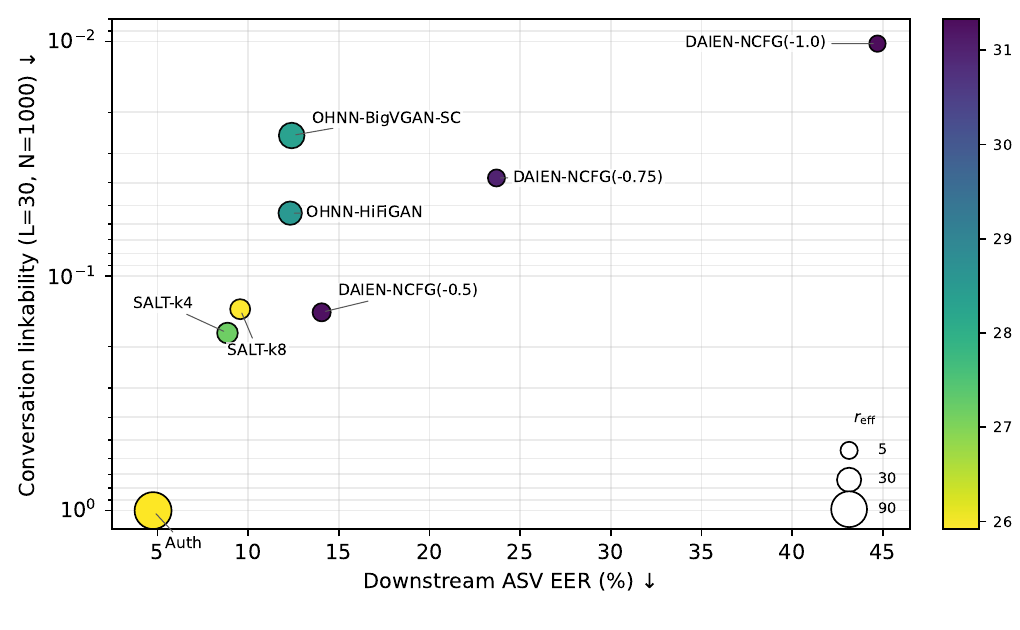}
\textbf{(b)} Conversation-level linkability
\caption{ Privacy--utility--diversity frontiers. The horizontal axis is downstream ASV EER (lower is better). Panel (a) uses external-verifier unlinkability EER, and panel (b) uses top-1 conversation-level linkability for 30 utterances per speaker and 1,000 speakers on an inverted logarithmic axis; higher positions indicate stronger privacy. Circle area denotes $r_{\mathrm{eff}}$, and color denotes pWER.}
\label{fig:frontiers}
\end{figure}

Fig.~\ref{fig:frontiers} visualizes the tabulated results. The horizontal axis shows the downstream performance of ASV measured by EER, reflecting how useful the corpus is. The vertical axis shows the degree of privacy protection. The color of each circle encodes the pseudo WER, and the size of the circle represents the diversity of speakers in the corpus, quantified by $r_{\mathrm{eff}}$. Panel (a) shows utterance-level unlinkability, whereas panel (b) uses the linkability at the top-1 conversation-level after aggregating 30 utterances from each speaker. Here, the linkability axis in panel (b) is inverted so that stronger privacy appears upward in both panels. 

We first observe that Auth lies in the region with high utility and high diversity but low privacy, as anticipated. DAIEN-NCFG($-1.0$) lies at the opposite end of the spectrum: it offers the strongest privacy protection but substantially harms downstream ASV performance. Panel (a) further shows that SALT-k4 and SALT-k8 remain in the highest-utility area among the anonymized datasets, while OHNN-BigVGAN-SC maintains the widest speaker distribution.

Comparing the two panels, OHNN-BigVGAN-SC exhibits consistent privacy behavior at both evaluation levels. Its unlinkability EER is 38.02\%, exceeding both OHNN-HiFiGAN and SALT-based methods. After utterance aggregation, its linkability is below all the other methods except for the extreme DAIEN setting. OHNN-BigVGAN-SC provides stronger privacy with comparable downstream utility, while the SALT-based methods show the best downstream ASV performance.

\begin{table*}[htbp]
\caption{Main results. Unlinkability and ASV utility are EER percentages. Legal-risk scores use $L=30$ aggregated utterances per speaker and population size $N=1000$. Speaker diversity is measured by mean pairwise cosine distance $\bar{d}$ and effective rank $r_{\mathrm{eff}}$. Gender and accent FDR are evaluated at the pooled-EER threshold. Arrows indicate the preferred direction.}
\begin{center}
\label{tab:main}
\begin{tabular}{lrrrrrrrrr}
\toprule
\multirow{2}{*}{Method} & \multirow{2}{*}{Unlink. EER $\uparrow$} & \multicolumn{2}{c}{Legal risk $\downarrow$} & \multicolumn{2}{c}{Utility $\downarrow$} & \multicolumn{2}{c}{Diversity $\uparrow$} & \multicolumn{2}{c}{FDR $\uparrow$} \\
\cmidrule(lr){3-4}\cmidrule(lr){5-6}\cmidrule(lr){7-8}\cmidrule(lr){9-10}
 & & Link. & S-out & ASV & pWER & $\bar{d}$ & $r_{\mathrm{eff}}$ & Gender & Accent \\
\midrule
Auth & 1.49 & 1.000 & 1.000 & 4.75 & -- & 0.928 & 97.25 & 0.945 & 0.930 \\
OHNN-HiFiGAN & 32.70 & 0.054 & 0.507 & 12.32 & 28.53 & 0.592 & 27.03 & 0.987 & 0.872 \\
OHNN-BigVGAN-SC & 38.02 & 0.025 & 0.507 & 12.40 & 28.31 & \textbf{0.673} & \textbf{36.27} & 0.986 & \textbf{0.895} \\
SALT-k4 & 26.84 & 0.175 & 0.664 & \textbf{8.86} & 27.17 & 0.257 & 16.50 & 0.967 & 0.772 \\
SALT-k8 & 26.08 & 0.138 & 0.684 & 9.56 & \textbf{25.92} & 0.250 & 14.15 & 0.963 & 0.798 \\
DAIEN-NCFG(-1.0) & \textbf{44.28} & \textbf{0.010} & \textbf{0.383} & 44.71 & 31.33 & 0.280 & 2.93 & \textbf{0.998} & 0.747 \\
DAIEN-NCFG(-0.75) & 39.97 & 0.038 & 0.436 & 23.70 & 30.99 & 0.333 & 4.80 & 0.698 & 0.854 \\
DAIEN-NCFG(-0.5) & 33.39 & 0.143 & 0.572 & 14.05 & 31.26 & 0.427 & 8.03 & 0.777 & 0.885 \\
\bottomrule
\end{tabular}

\end{center}
\end{table*}

\subsection{Linkability under Utterance Aggregation}

\begin{figure}[t]
\centering
\includegraphics[width=\linewidth]{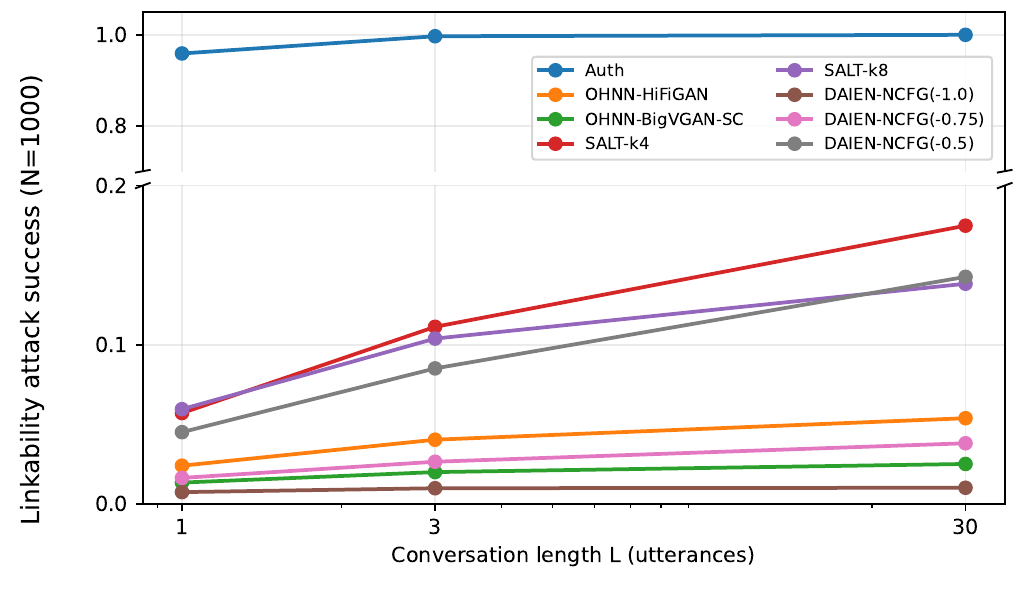}
\caption{ Linkability attack success as a function of the number of utterances aggregated per speaker for $N=1000$. The vertical axis omits the empty interval from 0.2 to 0.7.}
\label{fig:legal}
\end{figure}

Next, we analyze conversation-level privacy attacks. Fig.~\ref{fig:legal} illustrates how the success rate of the linkability attack varies with the number of aggregated utterances. As the number of utterances grows, the remaining identity-related information becomes increasingly detectable. With 30 utterances per speaker and 1000 speakers, the original VoxTube corpus is almost totally linkable, and all evaluated anonymization methods exhibit a similar pattern. OHNN-BigVGAN-SC exhibits the best conversation-level privacy except for DAIEN-NCFG($-1.0$), which implies the effectiveness of SC training.
The singling-out risk (``S-out’’ in Table~\ref{tab:main}) follows a similar trend as the conversation length increases. DAIEN-NCFG($-1.0$) is closest to chance for both linkability and singling-out, but again, this setting substantially degrades utility and speaker diversity.

\subsection{Linguistic Consistency}

SALT-k8 achieves the lowest pWER (25.92\%), followed by SALT-k4 (27.17\%), OHNN-BigVGAN-SC (28.31\%), and OHNN-HiFiGAN (28.53\%). The DAIEN-NCFG variants group around 31\%. SALT-k8 preserves linguistic information despite its reduced speaker diversity, while the two OHNN pipelines have comparable linguistic consistency.

\subsection{Subgroup Fairness}

In addition to the pooled-threshold FDR shown in Table~\ref{tab:main}, Table~\ref{tab:fairness} presents the absolute EERs for the female and male subgroups, as well as for the accent groups with the best and worst performance. In an ideal scenario, a fair and unbiased training dataset would yield comparable EERs across female and male subgroups and across the various accent subgroups.
\begin{table}[htbp]
\caption{Fairness summary on VoxCeleb2 gender and VoxAccent trials. EER is in percent and lower is better; accent codes are shown in parentheses. Pooled-threshold FDR is reported in Table~\ref{tab:main}.}
\begin{center}
\label{tab:fairness}
\footnotesize
\setlength{\tabcolsep}{3pt}
\begin{tabular}{lrrrr}
\toprule
\multirow{2}{*}{Method}
& \multicolumn{2}{c}{Gender EER}
& \multicolumn{2}{c}{Accent EER} \\
\cmidrule(lr){2-3}
\cmidrule(lr){4-5}
& F & M & Best & Worst \\
\midrule
Auth & 7.15 & 4.14 & 2.07 (FRA) & 15.38 (SPA) \\
OHNN-HiFiGAN & 15.04 & 11.08 & 7.08 (NET) & 18.72 (SPA) \\
OHNN-BigVGAN-SC & 13.27 & 10.51 & 7.08 (NET) & 18.63 (SPA) \\
SALT-k4 & 11.50 & 9.18 & 5.91 (NET) & 20.35 (SPA) \\
SALT-k8 & 12.30 & 9.55 & 6.25 (NET) & 20.21 (SPA) \\
DAIEN-NCFG(-1.0) & 44.69 & 45.86 & 38.04 (SPA) & 45.04 (UK) \\
DAIEN-NCFG(-0.75) & 26.66 & 19.11 & 10.52 (FRA) & 18.62 (NET) \\
DAIEN-NCFG(-0.5) & 16.81 & 11.73 & 4.81 (FRA) & 16.64 (SPA) \\
\bottomrule
\end{tabular}

\end{center}
\end{table}

Table~\ref{tab:fairness} indicates that female EER is higher than male EER in all conditions except for the collapsed DAIEN-NCFG($-1.0$). OHNN-BigVGAN-SC and OHNN-HiFiGAN exhibit the highest gender FDR among anonymized systems (0.986 and 0.987), but their subgroup EERs remain higher than those of Auth.
For accent, Spanish is the most challenging condition for most systems, suggesting that the anonymized corpora do not preserve accent variability uniformly.

OHNN-BigVGAN-SC maintains the highest accent FDR among anonymized systems (0.895), below Auth (0.930), while the SALT variants range between 0.772 and 0.798.
DAIEN-NCFG($-1.0$) exhibits gender EER values close to 45\% and accent EER values from 38.04\% to 45.04\%. Thus, the seemingly high gender FDR actually indicates an almost uniform failure rather than comparable effectiveness across groups. This underscores the importance of reporting pooled-threshold FDR alongside subgroup EER. These results show that certain subgroup disparities remain in VoxTubeS, and reducing them is an important direction for future work.

\subsection{Speaker Diversity Analysis}

The VoxTubeS corpora show pronounced differences in speaker diversity (Table~\ref{tab:main}). OHNN-BigVGAN-SC has $r_{\mathrm{eff}}=36.27$, substantially below Auth (97.25) but above other synthetic variants.

The behavior of $r_{\mathrm{eff}}$ broadly aligns with downstream ASV performance: when speaker diversity in the training corpus is substantially reduced, the resulting model loses variation needed to learn robust speaker embeddings. OHNN-BigVGAN-SC nevertheless shows that retaining more diversity than OHNN-HiFiGAN does not necessarily increase conversation-level linkability. The mean pairwise cosine distance $\bar{d}$ complements $r_{\mathrm{eff}}$ by measuring inter-speaker separation; OHNN-BigVGAN-SC is the highest anonymized variant on both measures, indicating broader but still substantially compressed speaker structure relative to Auth.

\subsection{Gender Structure after Anonymization}

\begin{figure}[t]
\centering
\includegraphics[width=\linewidth]{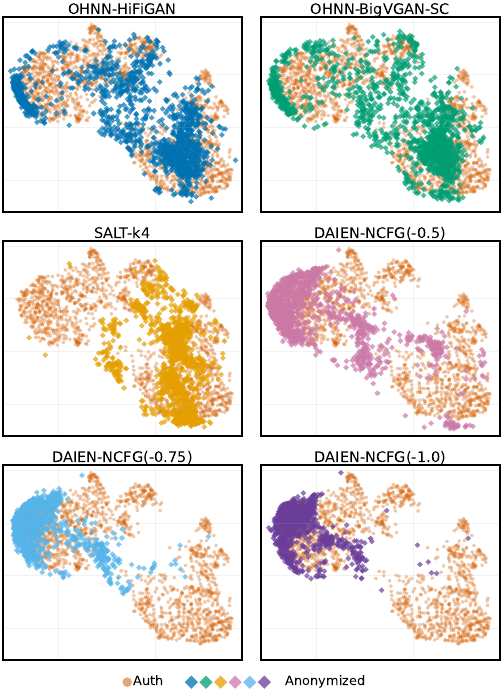}
\caption{UMAP of the speaker embedding centroids of 1,511 original speakers and the anonymized variants. The projection is fitted on authentic VoxTube centroids (red circles) and reused to transform each anonymized variant (diamonds). Official VoxTube metadata identify the two authentic clusters as predominantly male on the left and predominantly female on the right. SALT-k8 is omitted because its geometry is similar to that of SALT-k4.}
\label{fig:umap}
\end{figure}

Fig.~\ref{fig:umap} visualizes the authentic and anonymized speaker centroids. We computed one centroid per speaker for the authentic and anonymized utterances. UMAP \cite{mcinnes2018umap} is fitted once to authentic speaker centroids and then used to transform the centroids from each anonymized set, so displacement and concentration are comparable between panels. We used the official VoxTube language and gender metadata \cite{yakovlev2023voxtube,voxtubeMetadata}, which covers all 1,511 English speakers.

The projection forms two authentic clusters: the left cluster contains 800 male and 21 female speakers (97.4\% male), while the right cluster contains 659 female and 31 male speakers (95.5\% female). This agrees with prior evidence that speaker representations and similarity matrices can contain strong sex-related structure \cite{noe2023hiding}. OHNN-HiFiGAN and OHNN-BigVGAN-SC preserve the authentic cluster assignment for 87.6\% and 82.7\% of speakers, respectively, and OHNN-BigVGAN-SC covers slightly broader speaker space, consistent with its higher $r_{\mathrm{eff}}$. 
SALT-k4 moves 1,321 of 1,511 speakers toward the female-dominated cluster, while DAIEN-NCFG($-1.0$) places 1,504 speakers in the male-dominated lower-left region; the milder DAIEN-NCFG settings lie in between. Thus, anonymization not only changes individual identities but can also preserve, blur, or collapse gender-related structure in the speaker space.

\section{Discussion}

\begin{figure}[t]
\centering
\includegraphics[width=\linewidth]{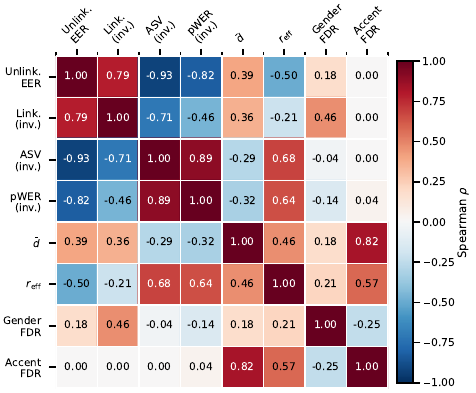}
\caption{ Spearman rank correlations among the seven generated variants, excluding Auth. Axes marked ``inv.'' are oriented so higher is preferred: Link. is one minus conversation linkability at $L=30,N=1000$, and ASV and pWER are negated error rates. Negative values indicate trade-offs and positive values indicate metrics improving together.}
\label{fig:metric_corr}
\end{figure}

The multi-axis evaluation reveals clear trade-offs among privacy, utility, and speaker diversity. Authentic speech remains highly linkable under ASV attack, whereas DAIEN-NCFG($-1.0$) achieves strong identity suppression by removing much of the variation needed for a useful training corpus. Utterance-level EER measures confusion on individual trials, while conversation-level linkability reveals speaker-specific cues that remain consistent across multiple utterances. Fig.~\ref{fig:metric_corr} summarizes the relationships across the seven generated variants. All metrics are oriented so that larger values are preferred; negative correlations thus indicate trade-offs.

Across the seven generated variants, Unlink. EER is negatively correlated with inversed ASV EER and inversed pWER. These two utility measures correlate with each other and are both moderately correlated with $r_{\mathrm{eff}}$. $\bar{d}$ correlates with accent FDR ($\rho=0.82$), suggesting that greater average inter-speaker separation may be associated with more balanced performance across accent groups. Gender FDR is nearly uncorrelated with utility and diversity.

The methods occupy distinct operating regions rather than forming a single ranking. OHNN-BigVGAN-SC provides stronger unlinkability and lower conversation-level linkability than OHNN-HiFiGAN, with nearly identical downstream ASV EER and pWER. It also retains more speaker diversity, although both OHNN variants remain substantially below Auth.

SALT provides the strongest utility. SALT-k4 achieves the best anonymized downstream EER, while SALT-k8 provides the best pWER and slightly lower long-conversation linkability. Both variants nevertheless show substantial diversity collapse and relatively weak accent FDR. DAIEN-NCFG exposes a tunable privacy control. Using more aggressive negative CFG improves privacy but sharply reduces downstream utility and speaker-space diversity. Thus, SALT is preferable when utility is central, whereas DAIEN-NCFG offers stronger identity suppression at the cost of corpus usefulness.

These results support different choices for different targets in real-world scenarios. OHNN-BigVGAN-SC offers the strongest combined privacy among the OHNN variants while maintaining similar utility and the broadest anonymized speaker space. SALT variants provide the strongest ASV utility, whereas DAIEN-NCFG($-1.0$) provides the highest privacy at a substantial cost to utility and diversity. Multiple DAIEN-NCFG operating points may therefore be useful when different privacy levels are required.

This study has several limitations. The privacy scores depend on the attacker model, its assumed knowledge, and the evaluation protocol, while the speaker-consistency objective and diversity measures all rely on ECAPA embeddings. The linguistic-consistency analysis also uses Whisper-generated transcripts rather than human references, and the same transcripts condition DAIEN-NCFG synthesis. Consequently, privacy and utility findings may vary with alternative speaker encoders, attacker assumptions, or transcription systems.

\section{Conclusion}

We introduced VoxTubeS, a family of speaker-anonymized synthetic speech corpora, and evaluated the variants with comprehensive analysis. The systems occupy a privacy--utility--diversity frontier: the strongest anonymizer can produce a poor dataset, while the most useful synthetic corpus can preserve residual linkability. 
Experiments also show that the speaker consistency objective can improve privacy while retaining comparable utility and moderately greater speaker diversity. Across all methods, no single system dominates every axis: SALT provides the strongest ASV utility, DAIEN-NCFG can provide stronger identity suppression at substantial utility and diversity cost, and subgroup fairness does not follow aggregate performance.
VoxTubeS therefore frames privacy-aware corpus construction as a choice among operating points with clearly documented trade-offs, requiring joint evaluation of utterance- and conversation-level privacy, downstream utility, linguistic consistency, speaker-space geometry, subgroup behavior, and responsible redistribution under the source license.

\section*{Acknowledgment}
This study is funded by MEXT KAKENHI Grants (24K21324). The experiments were carried out using the TSUBAME4.0 supercomputer at Institute of Science Tokyo. The authors thank Natalia Tomashenko, Xiaoxiao Miao, and Erica Cooper for their helpful comments on the manuscript.

OpenAI ChatGPT was used to assist in language editing and polishing of sections, limited to suggestions to improve grammar, clarity, wording, and typography. The authors reviewed and revised all suggestions and take full responsibility for the final content. The AI system was not used to generate research ideas, experimental data, results, figures, or conclusions.
\bibliographystyle{IEEEtran}
\bibliography{refs}

\end{document}